\documentclass[10pt,conference]{IEEEtran}
\usepackage[latin1]{inputenc}
\usepackage[numbers, square, comma, compress]{natbib}
\usepackage[english]{babel}
\usepackage{dsfont}
\usepackage{setspace}
\usepackage{amsfonts}
\usepackage{amssymb}
\usepackage{amsmath}
\usepackage{mathrsfs}
\usepackage{xcolor}
\usepackage{graphicx}
\usepackage{mathdots}
\usepackage{float}
\usepackage{dblfloatfix}
\usepackage{stmaryrd}
\usepackage{epstopdf}
\usepackage[colorinlistoftodos]{todonotes}
\usepackage[ruled,vlined]{algorithm2e}
\usepackage{verbatim}
\usepackage{tikz}
\usetikzlibrary{shapes,arrows}
\usetikzlibrary{calc}

\newtheorem{teo}{Theorem}

\newtheorem{lem}{Lemma}

\newtheorem{oss}{Remark}
\newtheorem{defi}{Definition}

\DeclareMathOperator{\tv}{\mathbb V}

\newcommand{\vv}[1]{``#1''}
\renewcommand{\IEEEQED}{\IEEEQEDopen}

\IEEEoverridecommandlockouts

\title{Strong Coordination of Signals and Actions over Noisy Channels}

\author{\IEEEauthorblockN{Giulia Cervia\IEEEauthorrefmark{1}, Laura Luzzi\IEEEauthorrefmark{1}, Ma\"{e}l Le Treust\IEEEauthorrefmark{1} and Matthieu R. Bloch\IEEEauthorrefmark{3}
\thanks{The work of M.R. Bloch was supported in part by NSF under grant CIF 1320304. The work of M. Le Treust was supported by INS2I CNRS through projects JCJC CoReDe 2015 and PEPS StrategicCoo 2016. This work was conducted as part of the project Labex MME-DII (ANR11-LBX-0023-01). }
}
\IEEEauthorblockA{\IEEEauthorrefmark{1} ETIS UMR 8051, Universit\'{e} Paris Seine, Universit\'{e} Cergy-Pontoise, ENSEA, CNRS, Cergy, France. \\
Email: \{giulia.cervia, laura.luzzi, mael.le-treust\}@ensea.fr}
\IEEEauthorblockA{\IEEEauthorrefmark{3}School of Electrical and Computer Engineering, Georgia Institute of Technology, Atlanta, Georgia\\
Email: matthieu.bloch@ece.gatech.edu} 
}

\begin{document}
\bstctlcite{IEEEexample:BSTcontrol}
\IEEEoverridecommandlockouts
\maketitle
\begin{abstract}
We develop a random binning scheme for strong coordination in a network of two nodes separated by a noisy channel, 
in which the input and output signals have to be coordinated with the source and its reconstruction. 
 In the case of non-causal encoding and decoding, we propose a joint
source-channel coding scheme and develop inner and outer bounds 
 for the strong coordination region. While the set of achievable
target distributions is the same as for empirical coordination, we characterize the
 rate of common randomness required for strong coordination.
\end{abstract}

\section{Introduction}
The 5G standard envisions direct device-to-device communication, 
which is likely to be a key enabler of the Internet of Things. In this decentralized network of connected objects,
such as wireless sensors, medical and wearable devices, smart energy meters, 
home appliances, and self-driving cars, devices will communicate with 
each other while sensing or acting on their environment.
It is essential that these devices, considered as autonomous decision-makers, 
cooperate and coordinate their actions. 

From an information theory perspective, two different metrics have been proposed to measure the level of coordination:
\emph{empirical coordination}, which requires the joint histogram of the actions to approach a target distribution, 
and \emph{strong coordination}, which requires the total variation distance of the distribution of sequences of actions 
to converge to an i.i.d. target distribution \cite{cuff2010}.
While empirical coordination investigates the average behavior over time, strong coordination is to be preferred from a security standpoint, 
since it guarantees that the sequence of actions will be unpredictable to an outside observer.
This is a consequence of the fact that statistical tests will produce identically distributed outcomes for distributions that are
close in total variation. 

Strong coordination with error free links has been studied in \cite{cuff2010} and the case 
in which only the source and the reconstruction have to be coordinated has been considered in \cite{haddadpour2013possible}.
However, in a realistic scenario where the communication links are noisy, the
signals that are transmitted and received over the physical channel 
become a part of what can be observed. 
One may therefore wish to coordinate both behaviors and communication \cite{cuff2011hybrid}. In this setting, strong coordination is desirable since the synthesized sequences would appear to be i.i.d. even from the perspective
of a malicious eavesdropper who can observe the signals sent over the communication channel  \cite{satpathy2016secure}.

In this paper, we address this problem in a two-node network comprised of an information source and 
a noisy channel, in which both nodes have access to a common source of randomness. 
An inner bound for the empirical coordination region has already been established in \cite{cuff2011hybrid} and
we focus here on the problem of achieving strong coordination for the same setting.  
This scenario presents two conflicting goals: 
the encoder needs to convey a 
message  to the decoder to coordinate the reconstructed version of the source,
while simultaneously coordinating the signals coding the message.
We derive an inner and an outer bound 
for the strong coordination region by developing a joint source-channel scheme in which an auxiliary codebook allows us 
to satisfy both goals. 
Since the two bounds do not match, 
the optimality of our scheme remains an open question. 
While the set of achievable target distributions is the same as for empirical coordination, 
we show that a positive rate of common randomness is required for strong coordination. 

The remainder of the paper is organized as follows. 
$\mbox{Section \ref{sec: prel}}$ introduces the notation, $\mbox{Section \ref{sec: sys}}$ 
describes the model under investigation and 
states the main result.
$\mbox{Section \ref{R1}}$ proves an inner bound by proposing 
a random binning scheme and a random coding scheme that have the same statistics. 
Finally, $\mbox{Section \ref{R2}}$ proves an outer bound.

\section{Preliminaries}\label{sec: prel}
We define the integer interval $[a,b]$ as the set of integers between $a$ and $b$.
Given a random vector $X^{n}:= (X_1, \ldots, X_n)$, we note $X^{i}$ the first $i$ components of $X^{n}$.
We note $\mathbb V (\cdot , \cdot)$ 
the variational distance 
between two distributions.

We now recall some useful results that we use later. 
\begin{lem}[Source coding with side information at the decoder]\label{lem1}
Consider an encoder that observes a sequence $X^n$ and transmits a message $M \in [1,2^{nR}]$ to a decoder that 
has access to side information $Y^n$, where $(X^n,Y^n)$ is a discrete memoryless source. If the encoding rate 
$R>H(X|Y)$, the decoder can recover $X^n$ from $M$ and $Y^n$ with arbitrarily small error probability.
\end{lem}

Lemma \ref{lem1} is a consequence of the Slepian-Wolf Theorem  \cite[Theorem 10.1]{elgamal2011nit}.

\begin{lem}\label{cor1}
Given a discrete memoryless source $(A^n,B^n)$ and $K= \varphi(B^n)$ a binning of $B^n$ with $2^{nR}$ values chosen 
independently and uniformly at random, 
if $R  < H(B|A)$, then we have
$$ \lim_{n \to \infty} \mathbb E_{\varphi} \left[\mathbb V \left( P_{A^n K}^{\varphi},  Q_K P_{A^n}\right)\right]=0,$$
where $\mathbb{E}_{\varphi}$ denotes the average over the random binnings, $P^{\varphi}$ is the distribution corrisponding to 
a fixed realization of the binning and $Q_K$ is the uniform distribution in $[1,2^{nR}]$. 
\end{lem} 

Lemma \ref{cor1} is a consequence of  \cite[Lemma 3.1]{ahlswede1998common} and \cite[Theorem 1]{yassaee2014achievability}.

\begin{oss}\label{cuff1617}
 We have,
 \begin{align*}
 & \tv (P_{A}, \widehat P_{A}) \leq \tv (P_{AB}, \widehat P_{AB}), \stepcounter{equation}\tag{\theequation}\label{cuff16}\\
 & \tv (P_A, \widehat P_A)= \tv (P_AP_{B|A}, \widehat P_A P_{B|A}), \stepcounter{equation}\tag{\theequation}\label{cuff17}
 \end{align*}
 where  \eqref{cuff16} and \eqref{cuff17} have been proven in \cite[Lemma 16]{cuff2009thesis} and \cite[Lemma 17]{cuff2009thesis} respectively.
\end{oss}

\vspace{-0.4cm}

\section{System model and main result}\label{sec: sys}
\vspace{-0.4cm}
 \begin{center}
\begin{figure}[h]
 \centering
 \includegraphics[scale=0.2]{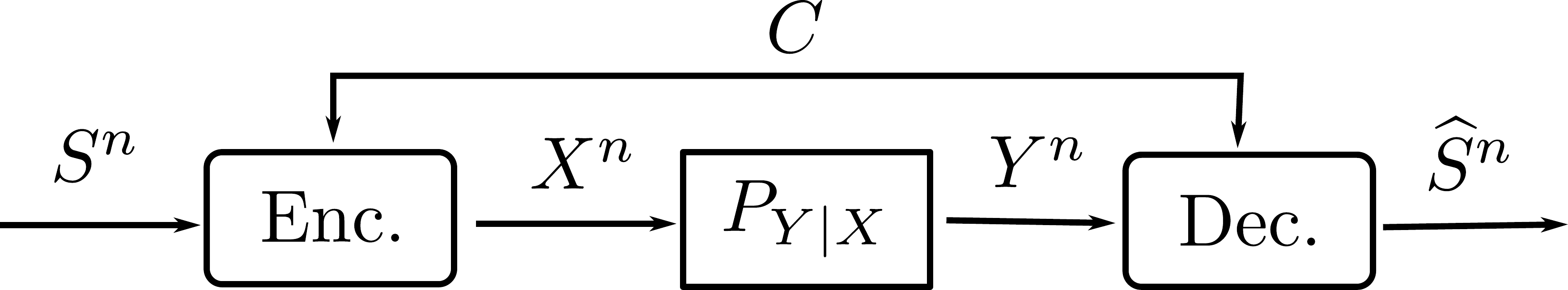}
 \vspace{-0.1cm}
\caption{Coordination of signals and actions for a two-node network with a noisy channel.}
\label{fig: coord}
\end{figure}
\end{center}
\vspace{-0.9cm}
Consider the model depicted in Figure \ref{fig: coord} in which two agents,
the encoder and the decoder, have access to a shared source of uniform randomness $C \in [1,2^{nR_0}]$. 
The encoder observes an i.i.d. source  $S^{n} \in \mathcal S^n$ with
distribution $\bar P_S$. The encoder then selects a signal $X^{n}= f_n(S^{n}, C)$,  $f_n: \mathcal S^n \times  [1,2^{nR_0}] \rightarrow \mathcal X^n$.
The signal $X^{n}$ is transmitted over a discrete memoryless channel parametrized by the conditional distribution $\bar P_{Y |X} $.
Upon  observing  $Y^{n}$ and $C$, the stochastic decoder selects an action $\widehat S^{n} = g_n(Y^{n}, C)$, $g_n: \mathcal Y^n  \times  [1,2^{nR_0}] \rightarrow \widehat{\mathcal S}^n$. For block length $n$, the pair $(f_n , g_n )$  constitutes a code. 
We recall the notions of achievability and the strong coordination region \cite{cuff2009thesis}.
\begin{defi}
A pair $(\bar{P}_{SXY\hat{S}}, R_0)$ is \emph{achievable} if there exists a 
sequence $(f_n,g_n)$ of encoders-decoders with rate of common randomness $R_0$, such that the induced 
joint distribution $P_{S^n X^nY^n \widehat S^n}$ is nearly indistinguishable 
from the i.i.d. distribution $\bar{P}_{SXY\hat{S}}$, in total variational distance:
\begin{equation*}
\lim_{n \to \infty} \tv \left(P_{S^n X^nY^n \widehat S^n}, \bar P_{SXY\widehat S}^{\otimes n} \right)=0.
\end{equation*}
The \emph{strong coordination region} $\mathcal{R}$ is the set of achievable pairs $(\bar P_{SXY\widehat S}, R_0)$.
\end{defi}

In the case of non-causal encoder and decoder, the problem of characterizing the strong
coordination region is still open, but we establish the following inner and outer bounds.
\begin{teo} \label{teouv}
Let $\bar P_{S}$ and $\bar P_{Y|X}$ be the given source and channel parameters, then 
$\mathcal R_1 \subseteq \mathcal{R} \subseteq \mathcal R_2$ where:
\begin{equation}\label{eq: region}
\mathcal R_1 := \begin{Bmatrix}
(\bar P_{SXY\widehat{S}}, R_0) \mbox{ } :\\
\bar P_{SXY\widehat{S}}= \bar P_{S} \bar P_{X|S} \bar P_{Y|X} \bar P_{\widehat{S}|SXY}  \\
\mbox{ }\exists \mbox{ } U \mbox{ taking values in $\mathcal U$}\\ 
\bar P_{SXYU \widehat S}= \bar P_{S} \bar P_{U|S} \bar P_{X|US} \bar P_{Y|X} \bar P_{\widehat{S}|UY}\\
\mbox{ } I(U;S) < I(U;Y)\\
\mbox{ } R_0 > I(U;SX\hat S|Y)\\
\mbox{ } \lvert \mathcal U \rvert \leq \lvert \mathcal S \rvert \lvert \mathcal X \rvert \lvert \mathcal Y \rvert  \lvert \widehat{\mathcal S} \rvert+1 \\
\end{Bmatrix}
\end{equation}
\begin{equation}\label{eq: region}
\mathcal R_2 := \begin{Bmatrix}
(\bar P_{SXY\widehat{S}}, R_0) \mbox{ } :\\
\bar P_{SXY\widehat{S}}= \bar P_{S} \bar P_{X|S} \bar P_{Y|X} \bar P_{\widehat{S}|SXY}  \\
\mbox{ }\exists \mbox{ } U \mbox{ taking values in $\mathcal U$}\\ 
\bar P_{SXYU \widehat S}= \bar P_{S} \bar P_{U|S} \bar P_{X|US} \bar P_{Y|X} \bar P_{\widehat{S}|UY}\\
\mbox{ } I(U;S) \leq I(X;Y)\\
\mbox{ } R_0 \geq I(U;SX\hat S|Y)\\
\mbox{ } \lvert \mathcal U \rvert \leq \lvert \mathcal S \rvert \lvert \mathcal X \rvert \lvert \mathcal Y \rvert  \lvert \widehat{\mathcal S} \rvert+1 \\
\end{Bmatrix}.
\end{equation}
\end{teo}

\begin{oss}
Even for empirical coordination, the problem of characterizing the coordination region is still open \cite{cuff2011hybrid}. 
The information constraint $I(U;S) \leq I(U;Y)$ for empirical coordination \cite[Theorem 1]{cuff2011hybrid}
is very similar to ours, as well as the decomposition of the joint probability distribution $\bar P_{S} \bar P_{U|S} \bar P_{X|US} \bar P_{Y|X} \bar P_{\widehat{S}|UY}$. 
The main difference is that strong coordination requires a positive rate of common randomness $R_0 \geq I(U;SX\hat S|Y)$.
\end{oss}

\begin{oss}
 Our inner bound is a generalization of the one in \cite{haddadpour2013possible} and the proof follows the 
 same strategy inspired by \cite{yassaee2014achievability}.
\end{oss}

\section{Proof of Theorem \ref{teouv}: inner bound}\label{R1}
First, we define two random schemes each of
which induces a joint distribution.
\vspace{-0.2cm}
\subsection{Random binning scheme}\label{rb2}
Assume that the sequences $S^n$ , $X^n$, $U^n$, $Y^n$ and $\widehat S^n$ are jointly i.i.d. with distribution $\bar P_{S^n} \bar P_{U^n| S^n} \bar P_{X^n| U^n S^n} \bar P_{Y^n|X^n} \bar P_{\widehat S^n|U^n Y^n}$. 
We consider two uniform random binnings for $U^n$:
\begin{itemize}
\item first binning $C = \varphi_1(U^n)$, where  $\varphi_1: \mathcal{U}^n \to [1,2^{nR_0}]$ maps each sequence of $\mathcal{U}^n$ uniformly and independently to the set $[1,2^{nR_0}]$;
\item second binning $F = \varphi_2(U^n)$,  $\varphi_2: \mathcal{U}^n \to [1,2^{n \tilde R}]$.
\end{itemize}
\vspace{-0.4cm}
 \begin{center}
\begin{figure}[b]
 \centering
 \includegraphics[scale=0.2]{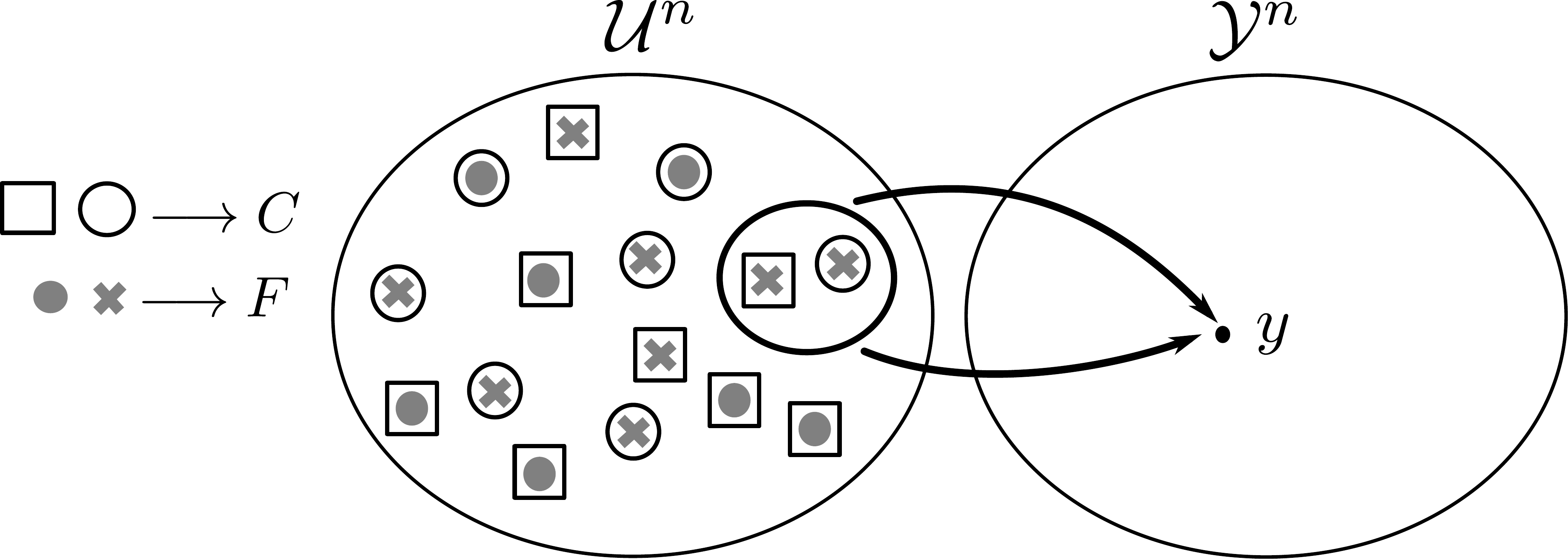}
 \vspace{-0.1cm}
\caption{The square and the circle represent the outputs of the first binning $C$ and the dot and the cross the outputs of the second binning $F$. Given $\mathbf y$ and the realizations of $C$ and $F$, it is possible to recover $\mathbf u$. }
\label{fig: db}
\end{figure}
\end{center}
\vspace{-0.4cm}
Note that if $\tilde R+R_0 >H(U|Y)$, by Lemma \ref{lem1}, it is possible to recover $U^n$ from $Y^n$ and $(C, F)$ with 
high probability using a Slepian-Wolf decoder via the conditional distribution $P^{SW}_{\widehat U^n|CF Y^n}$ as depicted in Figure \ref{fig: db}. 
This defines a joint distribution:
\begin{align*}
&\bar P_{S^n U^n \widehat U^n X^n Y^n C F \widehat S^n}=\\ 
  &  \bar P_{S^n}\!\bar P_{U^n|S^n} \bar P_{X^n|U^n \!S^n}  \bar P_{C|U^n} \bar P_{F|U^n}  \bar P_{Y^n|X^n} \!\bar P_{\widehat S^n|U^n\! Y^n} \!P^{SW}_{\widehat U^n|C\!F\! Y^n}\!.
\end{align*}
In particular, $\bar P_{U^n|CFS^n}$ is well defined.
\vspace{-0.2cm}
\subsection{Random coding scheme}\label{rc2}
\vspace{-0.2cm}
In this section we follow the approach in \cite[Section IV.E]{yassaee2014achievability} and \cite{haddadpour2013possible}.
Suppose that the encoder and decoder have access not only to common randomness $C$ 
but also to extra randomness $F$, where $C$ is generated uniformly at random 
in $[1, 2^{nR_0}]$ with distribution $Q_C$ and $F$ is generated uniformly at
random in $[1, 2^{n \tilde R}]$ with distribution $Q_F$ independently of $C$. 
Then the encoder generates $U^n$ according to $\bar P_{U^n|CFS^n}$ defined in Section \ref{rb2} and $X^n$ according to $\bar P_{X^n|S^n U^n}$.
The encoder sends $X^n$ through the channel.
The decoder gets $Y^n$ and $(C,F)$ and reconstructs $U^n$ 
via the conditional distribution  $P^{SW}_{\widehat U^n|CF Y^n}$. 
The decoder then generates $\widehat S^n$ letter by letter according to the distribution $P_{\widehat S^n|\widehat U^n Y^n}$
(more precisely $\bar P_{\widehat S^n| U^n Y^n}(\widehat{\mathbf s}|\widehat{\mathbf u}, \mathbf y)$, where $\widehat{\mathbf u}$ is the output of the Slepian-Wolf decoder).
This defines a joint distribution:
{\allowdisplaybreaks
\begin{align*}
& P_{S^n U^n \widehat U^n X^n Y^n C F \widehat S^n} =\\
 &  Q_C Q_F P_{S^n} \bar P_{ U^n|CFS^n} \bar P_{X^n|U^n S^n}\! \bar P_{Y^n|X^n} P^{SW}_{\widehat U^n|CF Y^n} P_{\widehat S^n|\widehat U^n Y^n}.
\end{align*}
}
We want to show that the distribution $\bar P$ is achievable  for strong coordination:
\begin{equation}\label{shat}
\lim_{n \to \infty} \tv \left(\bar P_{S^n  X^n U^n \widehat U^n Y^n \widehat S^n},  P_{S^n  X^n U^n \widehat U^n Y^n \widehat S^n}\right)=0.
\end{equation}
We prove that the random coding scheme possesses all the properties of the initial source coding 
scheme stated in Section \ref{rb2}.
Note that
\vspace{-0.2cm}
\begin{align*}
& \mathbb V (\bar P_{S^n U^n \widehat U^n X^n Y^n CF}, P_{S^n U^n \widehat U^n X^n Y^n C F}) \stepcounter{equation}\tag{\theequation}\label{chain1} \\
& = \mathbb V (\bar P_{S^n} \bar P_{U^n|S^n} \bar P_{X^n|U^n S^n} \bar P_{C|U^n}  \bar P_{F|U^n} \bar P_{Y^n|X^n} P^{SW}_{\widehat U^n|CF Y^n}, \\
& \phantom{= \tv |} Q_C Q_F P_{S^n} \bar P_{ U^n|CFS^n} \bar P_{X^n|U^n S^n} \bar P_{Y^n|X^n} P^{SW}_{\widehat U^n|CF Y^n}) \\
& {\overset{{(a)}}{=}}  \mathbb V ( \bar P_{S^n} \bar P_{U^n|S^n} \bar P_{C|U^n} \bar P_{F|U^n},   Q_C Q_F P_{S^n} \bar P_{U^n|CFS^n} ) \\
&{\overset{{(b)}}{=}} \mathbb V (\bar P_{S^ nCF},P_{S^n}  Q_C Q_F )
\end{align*} where $(a)$ and $(b)$ come from \eqref{cuff17}. 
Then if $R_0 + \widetilde R < H(U|S)$, we can apply Lemma \ref{cor1} where $B^n=U^n$,  $K=(C,F)$, $\varphi= (\varphi_1, \varphi_2)$, $A^n= S^n$ 
and find that
$$ \lim_{n \to \infty} \mathbb E_{\varphi} \left[\mathbb V \left( \bar P_{S^n CF}^{\varphi}, Q_C Q_F \bar P_{S^n}\right)\right]=0.$$
Therefore there exists a fixed binning $\varphi'$ such that, if we denote with  $\bar P^{\varphi'}$ and $P^{\varphi'}$  the distributions $\bar P$ and $P$ with respect to the choice of a binning $\varphi'$, we have
\begin{align*}
 \lim_{n \to \infty}  \mathbb V \left(\bar P_{S^nCF}^{\varphi'},P_{S^n}Q_C Q_F  \right)=0
\end{align*}
which by \eqref{chain1} implies 
\begin{align}\label{suxy}
 \lim_{n \to \infty} \mathbb V (\bar P^{\varphi'}_{S^n U^n \widehat U^n X^n Y^n CF}, P^{\varphi'}_{S^n U^n \widehat U^n X^n Y^n C F}) =0.
\end{align}
From now on, we will omit  ${\varphi'}$ to simplify the notation.

Now we would like to show that we have strong coordination for $\widehat S^n$ as well, but
in the second scheme $\widehat S^n$ is generated using $\widehat U^n$ and not $U^n$ as in the first scheme. Because of Lemma \ref{lem1}, the inequality $\widetilde{R}+R_0>H(U|Y)$  implies that $\widehat U^n$ is equal to $U^n$ with high probability and we will use this fact to show that
the distributions are close in total variational distance. 
First, we need to establish a technical lemma, whose proof can be found in the Appendix. 
\begin{lem}\label{nequ}
Let $V^n$ and $\widehat V^n$ such that 
$ \mathbb P \{\widehat V^n \neq V^n \} \to 0$ when $n \to \infty$.
Then for any random variable $W^n$ and for any joint distribution $P_{W^n V^n \widehat V^n}$ we have:
{\allowdisplaybreaks
\begin{align*}
& \lim_{n \to \infty} \tv(P_{W^n V^n \widehat V^n}, P_{W^n V^n} \mathds 1_{\widehat V^n| V^n})=0 \\
& \mbox{where } \mathds 1_{\widehat V^n| V^n} (\mathbf v| \mathbf v') = \begin{cases}
1 \quad \mbox{if } \mathbf v= \mathbf v'\\
0 \quad \mbox{if } \mathbf v \neq \mathbf v'
\end{cases}.
\end{align*}
}
\end{lem}
\vspace{-0.2cm}
Since $\widehat U^n$ is equal to $U^n$ with high probability, we can apply Lemma \ref{nequ} and if we denote $Z^n:=S^n X^n CF$ we find:
{\allowdisplaybreaks
\begin{align*}
& \lim_{n \to \infty} \tv(\bar P_{Z^n Y^n U^n \widehat U^n}, \bar P_{Z^nY^n U^n} \mathds 1_{\widehat U^n| U^n})=0, \stepcounter{equation}\tag{\theequation}\label{34yas'}\\
& \lim_{n \to \infty} \tv(P_{Z^n Y^n U^n \widehat U^n},P_{Z^n Y^n U^n} \mathds 1_{\widehat U^n| U^n})=0. \stepcounter{equation}\tag{\theequation}\label{35yas'}
\end{align*}
}
\vspace{-0.2cm}
Then using the triangle inequality, we find that 
{\allowdisplaybreaks
\begin{align*}
& \tv (\bar P_{Z^n Y^n U^n \widehat U^n \widehat S^n},P_{Z^n Y^n U^n \widehat U^n \widehat S^n} ) \\
& \!\!= \tv (\bar P_{Z^n Y^n U^n \widehat U^n} \bar P_{\widehat S^n | U^n Y^n}, P_{Z^n Y^nU^n \widehat U^n} P_{\widehat S^n | \widehat U^n Y^n} ) \stepcounter{equation}\tag{\theequation}\label{triu}\\
&\!\! \leq  \tv (\bar P_{Z^n Y^n U^n \widehat U^n}  \bar P_{\widehat S^n | U^n Y^n} , \bar P_{Z^n Y^n U^n} \mathds 1_{\widehat U^n| U^n} \bar P_{\widehat S^n | U^n Y^n} ) \\
&\!\!+\! \tv (\bar P_{Z^n Y^n U^n} \! \mathds 1_{\widehat U^n| U^n} \! \bar P_{\widehat S^n | U^n Y^n} ,\!  P_{Z^n Y^n U^n} \! \mathds 1_{\widehat U^n| U^n} P_{\widehat S^n \!| \widehat U^n Y^n} \!)\\
&\!\! +\! \tv (P_{Z^n Y^n U^n} \mathds 1_{\widehat U^n| U^n}  P_{\widehat S^n | \widehat U^n Y^n}, P_{Z^n Y^n U^n \widehat U^n} P_{\widehat S^n | \widehat U^n Y^n}) .
\end{align*}
}
The first and the third term go to zero by applying \eqref{cuff17} to \eqref{34yas'} and \eqref{35yas'} respectively.
Now observe that 
$ \mathds 1_{\widehat U^n| U^n} \bar P_{\widehat S^n | U^n Y^n}= \mathds 1_{\widehat U^n| U^n}  P_{\widehat S^n | \widehat U^n Y^n}$ 
by definition of $P_{\widehat S^n | \widehat U^n Y^n}$. 
Then by using \eqref{cuff17} again the second term 
is equal to $\tv \left(\bar P_{Z^n Y^n U^n}, P_{Z^n Y^n U^n}\right)$ that 
goes to zero by \eqref{suxy} and \eqref{cuff16}. 
Hence we have
\begin{equation}\label{zuys}
 \lim_{n \to \infty} \mathbb V (\bar P_{Z^n U^n \widehat U^n  Y^n \widehat S^n}, P_{Z^n U^n \widehat U^n Y^n \widehat S^n}) =0.
\end{equation}
\vspace{-0.2cm}
 Then by using \eqref{cuff16} we have proved \eqref{shat}.
\subsection{Remove the extra randomness F}\label{rf2}
\vspace{-0.2cm}
Even though the extra common randomness $F$ is required to coordinate $(S^n, {X}^n,Y^n,\widehat{S}^n,U^n)$, 
we will show that we do not need it in order to coordinate only $(S^n, {X}^n,Y^n,\widehat{S}^n)$.
Observe that by applying \eqref{cuff16}, equation \eqref{zuys} implies that 
\begin{equation}\label{convf2}
 \lim_{n \to \infty} \mathbb V (\bar P_{S^n  X^n Y^n  \widehat S^n F}, P_{S^n X^n Y^n  \widehat S^n F}) =0.
\end{equation}
As in \cite{yassaee2014achievability}, we would like to reduce the amount of 
common randomness by having the two nodes to agree on an instance $F=f$.
To do so, we apply Lemma \ref{cor1} again where $B^n=U^n$,  
$K=F$, $\varphi= \varphi''_2$ and $A^n= S^n X^n Y^n \widehat S^n$.
If $\tilde R < H(U| SXY \widehat S)$, there exists a fixed binning such that
\begin{align}\label{bin3}
 \lim_{n \to \infty} \tv \left(\bar P_{S^n X^n Y^n \widehat S^n F}, Q_F \bar P_{S^n X^n Y^n \widehat S^n}\right)=0.
\end{align}
\begin{oss}\label{binnings2}
Note that in Section \ref{rc2} we had already chosen a specific binning $\varphi'_2$. In the Appendix we prove that there exists a binning which works for both conditions.
\end{oss}
Because of \eqref{convf2}, \eqref{bin3} implies
\begin{equation}\label{bin4}
 \lim_{n \to \infty} \tv \left(P_{S^n X^n Y^n \widehat S^n F}, Q_F \bar P_{S^n X^n Y^n \widehat S^n}\right)=0.
\end{equation}
Hence, we can fix $f \in F$ such that $(S^n,X^n,Y^n, \hat S^n)$ is almost independent of $F$ according to $P$.
To conclude, we need  the following result proved in \cite[Lemma 4]{yassaee2014achievability}.
\begin{lem}\label{lem4}
If $\lim_{n \to \infty} \tv \left(P_{Y^n} P_{X^n|Y^n},P'_{Y^n} P'_{X^n |Y^n} \right) = 0$  then there exists $\mathbf y \in Y^n$ such that
 \begin{equation*}
 \lim_{n \to \infty} \tv \left(P_{X^n|Y^n= \mathbf y},P'_{X^n |Y^n= \mathbf y} \right) = 0.
 \end{equation*}
\end{lem}

If $f \in F$ is fixed, the distribution $P_{ S^n X^n Y^n \widehat S^n}$ changes to $P_{S^n X^n Y^n \widehat S^n|F=f}$
and by  Lemma \ref{lem4} we have
\begin{equation*}
 \lim_{n \to \infty} \mathbb V (\bar P_{S^n X^n Y^n \widehat S^n|F=f}, P_{S^n X^n Y^n \widehat S^n|F=f}) =0.
\end{equation*}
Since $\bar P_{S^n X^n Y^n \widehat S^n|F=f}$ is close to $\bar P_{S^n X^nY^n\widehat{S}^n}$ because of \eqref{bin3}, we have
\begin{equation*}
 \lim_{n \to \infty} \mathbb V (\bar P_{S^n X^n Y^n \widehat S^n}, P_{S^n X^n Y^n \widehat S^n}) =0.
\end{equation*}

\vspace{-0.3cm}
\subsection{Rate constraints}
We have imposed the following rate constraints:
{\allowdisplaybreaks
\begin{align*}
H(U|Y) &< \widetilde R+R_0 < H(U |S)\\
 \widetilde R & < H(U|SXY \widehat S).
\end{align*}
}
Therefore we obtain:
{\allowdisplaybreaks
\begin{align*}
 & R_0 > H(U|Y)-H(U|SXY \widehat S) = I(U;SX\hat S|Y)\\
 & I(U;S)  < I(U;Y). \tag*{\IEEEQED} 
\end{align*}
}
\vspace{-0.3cm}
\section{Proof of Theorem \ref{teouv}: outer bound}\label{R2}
Consider a code $(f_n,g_n)$ that induces a distribution $P_{S^n X^n Y^n \widehat S^n}$ that is $\varepsilon$-close in $L^1$ distance to the i.i.d. distribution $\bar P_{S X Y \widehat S}^{\otimes n}$.
Let the random variable $T$ be uniformly distributed over the
set $[1,n]$ and independent of the induced joint distribution
$P_{S^n X^n Y^n \widehat S^n C}$. The variable $T$ will serve as a random time index. The variable $S_T$ is independent of $T$ because $S^n$ is an i.i.d. source sequence \cite{cuff2010}.
Then we have
\begin{align*}
& 0  \overset{(a)}{\leq} I(X^n;Y^n)-I(C,S^n;Y^n) \\ 
& \leq I(C,X^n;Y^n)-I(C,S^n;Y^n)\\
& = I(X^n;Y^n|C)-I(S^n;Y^n|C)+I(C;Y^n)-I(C;Y^n)\\
& = H(Y^n|C)-H(Y^n|X^nC)+H(S^n|Y^n C)-H(S^n|C)\\
& \overset{(b)}{\leq} \sum_{t=1}^n \left(H(Y_t) \!-\!H(Y_t|X_t)\!+ \!H(S_t|S^{t-1} Y_t Y_{\sim t} C)\!- \!H(S_t)\right)\\
& \overset{(c)}{\leq} \sum_{t=1}^n \left(H(Y_t) -H(Y_t|X_t)+ H(S_t|Y_{\sim t} C)- H(S_t)\right)\\
& \overset{(d)}{\leq}\! n H\!(Y_T) \!-\!n\! H(Y_T|X_T,T)\! \!+ \!n\!H(S_T|Y_{\sim T} C T)\!-\!n\! H(S_T|T) \\
& \overset{(e)}{=} n H(Y_T) -n H(Y_T|X_T) + nH(S_T|Y_{\sim T} C T)-n H(S_T) \\
&= n I(X_T; Y_T)-n I( S_T;Y_{\sim T}, C, T )
\end{align*}
where $(a)$ comes from the Markov chain $Y^n-X^n-(C,S^n)$ and $(b)$ comes from the following facts: 
conditioning doesn't increase entropy,  $\bar P_{Y|X}$ is a memoryless channel, the chain rule for the conditional entropy and 
$S^n$ is an i.i.d. source independent of $C$.
Recall that we note $Y_{\sim t}$ the vector $(Y_i)_{i \neq t}$, $i\in [1,n]$, where the component $Y_t$ has been removed.
The inequalities $(c)$ and $(d)$ come from the fact that $H(Y_T|T)$ is smaller or equal to $H(Y_T)$ since conditioning doesn't increase entropy and 
$(e)$ from the memoryless channel $\bar P_{Y|X}$ and the i.i.d. source $\bar P_{S}$.

For the second part of the converse, we need to establish a technical result first. The proof is in the Appendix. 
\begin{lem}\label{lemmit}
Let $P_{X^n}$ such that 
$\tv\left(P_{X^n}, \bar P_{X}^{\otimes n}\right) \leq \varepsilon,$
then we have
\begin{equation*}
\sum_{t=1}^n I(X_t;X_{\sim t}) \leq n f(\varepsilon) 
\end{equation*}
where $f(\varepsilon)$ goes to zero as $\varepsilon$ does.
\end{lem}
Then we have
{\allowdisplaybreaks
\begin{align*}
& nR_0  \geq H(C) \geq H(C|Y^n) \geq   I(S^n X^n{\widehat S}^n; C|Y^n)\\
& = \sum_{t=1}^n I(S_t X_t{\widehat S}_t;C|S^{t-1} X^{t-1}{\widehat S}^{t-1} Y_{\sim t}Y_t)\\
&= \sum_{t=1}^n I(S_t X_t{\widehat S}_t;CS^{t-1} X^{t-1}{\widehat S}^{t-1}Y_{\sim t}|Y_t)\\
&- \sum_{t=1}^n I(S_tX_t{\widehat S}_t; S^{t-1} X^{t-1}{\widehat S}^{t-1} Y_{\sim t}|Y_t)\\
& \geq \sum_{t=1}^n  I(S_t X_t{\widehat S}_t;CY_{\sim t}|Y_t)\\
& - \sum_{t=1}^n I(S_t X_t{\widehat S}_t;S^{t-1} X^{t-1}{\widehat S}^{t-1} Y_{\sim t}|Y_t)\\
& \overset{(a)}{\geq} \sum_{t=1}^n I(S_t X_t{\widehat S}_t;CY_{\sim t}|Y_t)-n f(\varepsilon)\\
&=\!  n\!  I(S_T X_T{\widehat S}_T;CY_{\sim T}|Y_T T) -n f(\varepsilon) \\
& = \! n \! I(S_T X_T{\widehat S}_T;CY_{\sim T} T|Y_T )\! \! -n\!  I(S_T,X_T,{\widehat S}_T; T|Y_T) \!\! -\!\! n \! f(\varepsilon)\\
& \geq \!n I(S_T X_T{\widehat S}_T;CY_{\sim T} T|Y_T ) \!\! -n \! I(S_T,X_T,{\widehat S}_T, Y_T; T) \! \! - n\!  f(\varepsilon)\\
& \overset{(b)}{\geq} n I(S_T X_T{\widehat S}_T;CY_{\sim T} T|Y_T ) -2n f(\varepsilon)
\end{align*}
}
where $(a)$ follows from the following chain of inequalities
\begin{align*}
& \sum_{t=1}^n I(S_t X_t{\widehat S}_t;S^{t-1} X^{t-1}{\widehat S}^{t-1} Y_{\sim t}|Y_t)\\
\leq  &\sum_{t=1}^n I(S_t X_t{\widehat S}_t;S_{\sim t} X_{\sim t}{\widehat S}_{\sim t} Y_{\sim t}|Y_t)\\
 \leq &\sum_{t=1}^n I(S_t X_t{\widehat S}_t Y_t;S_{\sim t} X_{\sim t}{\widehat S}_{\sim t} Y_{\sim t}) \leq n f(\varepsilon)
\end{align*}
and $f(\varepsilon)$ is defined in Lemma \ref{lemmit}.
Finally, the proof of $(b)$ comes from \cite[Lemma VI.3]{cuff2013distributed}.\\
We identify the auxiliary random variables $U_t$ with $(C,Y_{\sim t})$ for each $t \in [1,n]$ and $U$ with
$(C,Y_{\sim T}, T)$. For each $t \in [1,n]$ the following two Markov chains hold: $(S_t,X_t)-(C,Y_{\sim t},Y_t)-\widehat{S}_t$ 
and $Y_t-X_t-(C, Y_{\sim t}, S_t)$. Since $U=U_t$ when $T=t$, we also have $(S,X)-(U,Y)-\widehat{S}$ and $Y-X-(U,S)$. 
The cardinality bound comes from \cite[Lemma VI.1]{cuff2013distributed}. 



\appendix 
\subsection{Proof of Lemma \ref{nequ}}
We denote the event that $\widehat V^n$ is equal to  $V^n$ with $\mathcal A : = \{ V^n= \widehat V^n \}$. We know that $\mathbb P \{\mathcal A\}$ tends to 1.
We can write the joint distribution $ P_{W^n  V^n \widehat V^n}$ as
\begin{equation*}
\mathbb P \left\{\mathcal A\right\} P_{W^n V^n \widehat V^n| \mathcal A} + \mathbb P \left\{\mathcal A^c\right\} P_{W^n V^n \widehat V^n| \mathcal A^c}.
\end{equation*}
Hence, we have
\begin{align*}
& \tv(P_{W^n V^n \widehat V^n}, P_{W^n V^n} \mathds 1_{\widehat V^n| V^n}) \leq  \mathbb P \left\{\mathcal A^c\right\} {\lVert   P_{W^n V^n \widehat V^n| \mathcal A^c} \rVert}_{L^1}\\
& + {\lVert \mathbb P \left\{\mathcal A\right\} P_{W^n V^n \widehat V^n| \mathcal A}-P_{W^n V^n} \mathds 1_{\widehat V^n| V^n} \rVert}_{L^1} 
\end{align*}
where the first term is equal to $\left(1-\mathbb P \left\{\mathcal A\right\}\right) P_{W^n V^n} \mathds 1_{\widehat V^n| V^n}$ and goes to 0 since $\mathbb P \left\{\mathcal A\right\}$ tends to 1 and the second term goes to 0 since $ \mathbb P \left\{\mathcal A^c\right\}$ does.

\vspace{-0.2cm}
\subsection{Proof of Remark \ref{binnings2}}
We want to prove that there exists a binning which works for both the conditions in Section \ref{rc2} and Section \ref{rf2}.
If we denote with $\mathbb E_{\varphi_1 \varphi_2 }$ and $\mathbb E_{\varphi_2}$ the  expected value with respect to the random binnings, 
for all $\varepsilon$, there exists $\bar n$  such that $\forall n \geq \bar n$
{\allowdisplaybreaks
\begin{align*}
& \mathbb E_{\varphi_1 \varphi_2 } \left[\tv \left(\bar P^{\varphi_1 \varphi_2}_{ S^n FC}, Q_{F} Q_{C}  \bar P_{S^n} \right)\right]  < \frac{\varepsilon}{2}\\
& \mathbb E_{ \varphi_2} \left[\tv \left(\bar P_{S^n X^n Y^n \widehat S^n F}^{ \varphi_2},  Q_{F} \bar P_{S^n X^n Y^n \widehat S^n} \right)\right] < \frac{\varepsilon}{2}
\end{align*}
}
which implies by Markov's inequality
{\allowdisplaybreaks
\begin{align*}
&\mathbb P_{\varphi_1 \varphi_2} \left\{ \tv \left(\bar P^{\varphi_1 \varphi_2}_{S^nFC}, Q_{F} Q_{C}  \bar P_{S^n}\right) <  \varepsilon \right\} > \frac{1}{2} \stepcounter{equation}\tag{\theequation}\label{min122}\\
& \mathbb P_{\varphi_2}\left\{\tv \left(\bar P_{S^n X^n Y^n \widehat S^n F}^{\varphi_2}, Q_{F} \bar P_{S^n X^n Y^n \widehat S^n} \right)  < \varepsilon \right\} > \frac{1}{2}.
\end{align*}
}

In Section \ref{rc2} and \ref{rf2} we have chosen the binnings  $(\varphi'_1, \varphi'_2)$  and $\varphi''_2$ respectively such that 
{\allowdisplaybreaks
\begin{align*}
&\lim_{n \to \infty} \tv \left(\bar P^{\varphi'_1 \varphi'_2}_{S^nFC}, Q_{F} Q_{C}  \bar P_{S^n}  \right)=0\\
& \lim_{n \to \infty} \tv \left(\bar P^{\varphi''_2}_{S^n X^n Y^n \widehat S^n F},  Q_{F} \bar P_{S^n X^n Y^n \widehat S^n}\right)=0.
\end{align*}
}
It follows from \eqref{min122} that the intersection of the two sets is non-empty, therefore there exists a binning 
$\varphi_2^{*}$ that satisfies both conditions.

\vspace{-0.2cm}
\subsection{Proof of Lemma \ref{lemmit}}
The following result has already been proved in \cite[Lemma 2.7]{csiszar2011information}.
\begin{lem}\label{csi2.7}
Let $P$ and $Q$  two distributions on $\mathcal X $  
such that 
$\tv(P,Q) = \varepsilon$ and $\varepsilon \leq 1/2$, then
\vspace{-0.3cm}
\begin{equation*}
 \lvert H(P)-H(Q) \rvert \leq  \varepsilon \log{\frac{\lvert \mathcal X \rvert}{\varepsilon}}.
\end{equation*}
\end{lem}
We also need this lemma proved in \cite[Lemma 3.2']{yassaee2014achievability}.
\begin{lem}\label{yas3.2'}
If $\tv(P_{X}P_{Y|X}, Q_{X}Q_{Y|X})\leq \varepsilon$ then
\vspace{-0.2cm}
\begin{equation*}
 \mathbb P\{x \in \mathcal X | \tv(P_{Y|X=x}, Q_{Y|X=x}) \leq \sqrt{\varepsilon}\} \geq 1-2 \sqrt{\varepsilon}.\\
\end{equation*}
\end{lem}

Now, consider the set
\begin{equation*}
\mathcal B:=\{ \mathbf x \in  \mathcal X^{n-1}| \tv(P_{X_t |X_{\sim t}=\mathbf x}, \bar P_{X}) \leq \varepsilon\}.
\end{equation*}
By Lemma \ref{yas3.2'}, $\mathbb  P \{\mathcal B\} \geq 1-2 \sqrt{\varepsilon}$. 
Observe that 
{\allowdisplaybreaks
\begin{align*}
& H(X)-H(X_t|X_{\sim t})\\
& =  H(X) -   \sum_{\mathbf x \in \mathcal X^{n-1}} P_{X_{\sim t}}(\mathbf x)  H(X_t|X_{\sim t}= \mathbf x) \\
& \leq \! \! \sum_{\mathbf x \in \mathcal X^{n-1}} \! \left(P_{X_{\sim t}}(\mathbf x) H(X) -   P_{X_{\sim t}}(\mathbf x)  H(X_t|X_{\sim t}= \mathbf x) \right)\\
& =  \sum_{\mathbf x \in \mathcal B} \left( P_{X_{\sim t}}(\mathbf x) H(X)- P_{X_{\sim t}}(\mathbf x) \lvert H(X_t|X_{\sim t}= \mathbf x)\right) \\
& +  \sum_{\mathbf x \in \mathcal B^c} \left( P_{X_{\sim t}}(\mathbf x) H(X)- P_{X_{\sim t}}(\mathbf x) \lvert H(X_t|X_{\sim t}= \mathbf x)\right).
\end{align*}
}
Hence by Lemma \ref{csi2.7} 
\begin{equation*}
\lvert H(X_t|X_{\sim t}=\mathbf x)-H(X) \rvert \leq \varepsilon \log{\frac{\lvert \mathcal X \rvert}{\varepsilon}}.
\end{equation*}
Let $\delta:=\varepsilon \log{\frac{\lvert \mathcal X \rvert}{\varepsilon}}$, then the first term is bounded by 
\begin{equation*}
\sum_{\mathbf x \in \mathcal B} P_{X_{\sim t}}(\mathbf x) \delta \leq \delta,
\end{equation*}
while the second term is smaller than 
{\allowdisplaybreaks
\begin{align*}
&\mathbb P \{\mathcal B^c\} \left(H(X_t)+H(X)\right) \leq 2 \sqrt{\varepsilon} \left(2 H(X) + \delta \right).
\end{align*}
}
Again, by Lemma \ref{csi2.7}, we have 
\begin{equation*}
\lvert H(X_t)-H(X) \rvert \leq \delta.
\end{equation*}
Finally, $I(X_t;X_{\sim t})= H(X_t)-H(X) + H(X)-H(X_t|X_{\sim t})$ is smaller than $f(\varepsilon)= 2 \sqrt{\varepsilon} (2H(X)+\delta)+ 2\delta$.\\

\vspace{-0.2cm}
\begin{footnotesize}
\bibliographystyle{IEEEtran}

\end{footnotesize}
\end{document}